# Tuning multiple Landau Quantization in Transition-Metal Dichalcogenide with Strain

Zihao Huang[1,2#], Guoyu Xian[1,2#], Xiangbo Xiao[1#], Xianghe Han[1,2], Guojian Qian[1,2], Chengmin Shen[1,2], Haitao Yang[1,2*], Hui Chen[1,2*], Banggui Liu[1,2*], Ziqiang Wang[3] and Hong-Jun Gao[1,2]

[1] Beijing National Center for Condensed Matter Physics and Institute of Physics, Chinese Academy of Sciences, Beijing 100190, PR China

[2] School of Physical Sciences, University of Chinese Academy of Sciences, Beijing 100190, PR China

[3] Department of Physics, Boston College, Chestnut Hill, MA 02467, USA

## ABSTRACT

Landau quantization associated with the quantized cyclotron motion of electrons under magnetic field provides the effective way to investigate topologically protected quantum states with entangled degrees of freedom and multiple quantum numbers. Here we report the cascade of Landau quantization in a strained type-II Dirac semimetal $NiTe_2$ with spectroscopic-imaging scanning tunneling microscopy. The uniform-height surfaces exhibit single-sequence Landau levels (LLs) at a magnetic field originating from the quantization of topological surface state (TSS) across the Fermi level. Strikingly, we reveal the multiple sequence of LLs in the strained surface regions where the rotation symmetry is broken. First-principles calculations demonstrate that the multiple LLs attest to the remarkable lifting of the valley degeneracy of TSS by the in-plane uniaxial or shear strains. Our findings pave a pathway to tune multiple degrees of freedom and quantum numbers of TMDs *via* strain engineering for practical applications such as high-frequency rectifiers, Josephson diode and valleytronics.

Lifting and controlling over the degeneracy of multiple degrees of freedom such as spin[1,2], valley[3,4], orbital[5], helicity[6] and pseudospin[7] is the central goal of functional devices and electronics including spintronic and valleytronics. Quantum materials hosting spin-momentum-locked topological surface states (TSSs) that span the bulk bandgap[8] are a promising candidate for such applications. Landau quantization, which is a fundamental phenomenon associated with the quantized cyclotron motion of electrons in a magnetic field[9,10], provides the effective way to investigate topologically protected quantum states with entangled degrees of freedom and multiple quantum numbers. To date, single-sequence Landau quantization of degenerate electron has been reported in graphene and other topological materials[11–13]. However, investigation of multiple-sequence Landau levels emerged from the lift of multiple degrees of freedom to understand the origin of the exotic properties of quantum materials remains experimentally rare[14] and to further controlling the topologically-protected quantum states[15,16] remains experimentally elusive.

Recently, the class of transition metal dichalcogenides (TMDs) has been identified to be a material system with rich topological properties, including Dirac points in the bulk bands characteristic of Dirac semimetals and band inversion induced helical TSSs[17,18] characteristic of strong topological insulators. The TMDs $NiTe_2$[19–22] and $PdTe_2$[23] exhibit such a dual character. For example, angle-resolved photoemission spectroscopy (ARPES) measurements[19] demonstrate the type-II Dirac point sitting about 0.1 eV above the Fermi level ($E_F$), which originates from the transition metal $d$ orbital states. In addition, helical TSSs are detected to cross the $E_F$, which are remarkably split into two branches of opposite helicity[19,23]. The latter has recently been utilized to achieve the field-induced Josephson diode effect[24]. Moreover, the TSSs produce rich multivalley Fermi surfaces with complex spin textures[23]. Therefore, the TMDs provide a platform to investigate topologically protected quantum states with entangled degrees of freedom and multiple quantum numbers, which remains largely unexplored.

Here, we study the Landau quantization of the TSSs in type-II Dirac semimetal $NiTe_2$ single crystals by using scanning tunneling microscopy/spectroscopy (STM/S). We observe multiple groups of Landau levels (LLs) in the TSSs at a low temperature of 420 mK and in high magnetic fields applied perpendicular to the surface. In the uniform-height region, one group of LLs are observed in the energy range where the surface states are dominant. The LLs correspond to the quantization of TSS band across $E_F$ resulting from the band inversion in the chalcogen $p$-orbital manifold. In contrast, we observe two and three groups of LLs in strained surface regions where the rotation symmetry is broken. Combined with the density functional theory (DFT) calculations, we attribute the multiple groups of LLs to the lifting of the valley degeneracy that produces the additional valley-polarization of TSSs, as the strain breaks the rotation symmetry and causes anisotropic shifting of the surface bands along distinct momentum directions.

NiTe$_2$, crystallized in a trigonal structure with the P$\bar{3}$m1 space group[22], is a van der Waals coupling material. The unit cell consists of one Ni atom and two Te atoms (Figure 1(a)). Basic structural measurements demonstrate the high quality of as-grown 1T-NiTe$_2$ crystal (Figure S1). The large-scale STM image (Figure 1(b)) shows a flat and clean Te-terminated surface of NiTe$_2$ with low defect density. The atomically-resolved STM image obtained at the Te-terminated surface of NiTe$_2$ (inset of Figure 1(b)) shows a hexagonal lattice with lattice constants *a*=*b*≈0.38 nm which are consistent with the stoichiometric NiTe$_2$ in the literatures where *a*=*b*=0.38551 nm, and *c*= 0.52660 nm[20,21].

To investigate the fine structure of the TSSs and their topological properties of NiTe$_2$, we perform the Landau-level spectroscopy of STM[11,25–27] at 420 mK (electronic temperature is about 620 mK, see Methods) and strong magnetic field perpendicular to the surface ($B_z$). In the d*I*/d*V* spectra of the flat area with uniform-height distributions (Figure 1(b)) and negligible lattice distortions (Figure S2(a-d)), several sharp peaks emerge around the E$_F$ at $B_z$ = 11 T, which are absent at 0 T (Figure 1(c)). With increasing $B_z$, all the peaks shift significantly toward higher energy (Figure 1(d)), indicating that the peaks result from the Landau quantization of electron-like pockets with the band minimum lying below the E$_F$.

We then determine the band dispersion of the Landau quantized electron band based on the semiclassical Lifshitz-Onsager relation $S_n = 2\pi e(n + \gamma)B/\hbar$ ($\gamma = \frac{1}{2} - \frac{\phi_B}{2\pi}$, $\phi_B$ is the Berry phase enclosed by the cyclotron orbit). We assign the first peak of LLs at 9 T to different integers for a given value of $\gamma$ (Figure S3, S4) and then analyze the dispersion using the obtained LL sequence for different $\gamma$ values[28] (Figure S5). As a result, when we assign the quantum number of the first LL to $n = 4$ and the phase factor $\gamma$=1/2 (Berry phase is zero), the LL peak energy $E_n$ and the calculated $k_n$ converge to an effective TSS band. In the energy range from -8 meV to 8 meV, the Fermi velocity of the band ($v_F$) is about 1.4×10$^5$ m/s with the band minimum at about -40.6 meV (Figure 1(e)).

The TSSs induced by the band inversion in the chalcogen *p*-orbital manifold are observed in the spin-polarized ARPES measurements across the E$_F$ along $\bar{\Gamma} - \bar{M}$ direction (Figure 1(f)), whereas the bulk Dirac cone is about 0.1 eV above the E$_F$[19,21]. Therefore, we attribute the emerging LL peaks in the applied magnetic field to the quantization of the TSS near the E$_F$ instead of the type-II bulk Dirac cones at around 0.1 eV and other surface states far away from the E$_F$. The Dirac cone of TSS near the Γ point shows hole-like dispersion, which is inconsistent with the STM observations. In addition, the zero-value Berry phase enclosed by the Landau orbits further excludes that the Landau quantization of Dirac cone at Γ point where the Berry phase should be nonzero. On the other hand, the fitted dispersions and band minimum value are consistent with the electron-like subbands of TSS near $E_F$ (highlighted by the light purple shades in Figure 1(f)). Therefore, we conclude that the Landau quantization of electron-like subbands of TSSs along the $\bar{M}-\bar{\Gamma}-\bar{M}$ direction near $E_F$ gives rise to the single-group LLs in density of states (DOS) (Figure 1(f)).

We note that only parts (n=4, … 10) of the LL sequence are observable in the energy range from -10 mV to 10 mV around the $E_F$. Moreover, the static peaks, namely the Landau levels with index $n = 0$, are absent in the $B_z$-dependent d$I$/d$V$ spectra (Figure 1(d)). We attribute the absence of the other LLs to the overlap of the surface states with the bulk band[12,29] outside this energy range. In addition, at the 0 T, the d$I$/d$V$ maps of the NiTe$_2$ surface at low energy clearly show the standing-wave patterns originated from the intervalley scattering of surface state (Figure S6). The energy range for observing the standing-wave patterns is around -10 mV to 10 mV (Figure S6(b),(c)), coinciding with the energy range of the observed LLs.

To further investigate the effects of strain[30,31] on the LLs, we measure the Landau-level spectroscopy at strained surfaces. In the strained region, the height distribution is not uniform (Figure 2(a)) and the lattice shows distortion (Figure S2(e-h)). Strikingly different from the single-series of LLs in graphene and topological insulators[12,32], there are two groups of LLs (labeled as *na* and *nb, n*=0,1,2…), indicating two different origins of the Landau-quantized states (Figure 2(b)). The two groups of LLs (highlighted by azure and orange shades in Figure 2(b)) show energy shifts with increasing $B_z$. At $B_z$ = 11 T, the d$I$/d$V$ spectra obtained along a relatively long distance in the region of Figure 2(a) show that the two groups of LLs are spatially homogeneous over the strain region, excluding the local effects that may cause multiple LLs at specific locations (Figure 2(e)). Furthermore, the Landau-level spectroscopy at $B_z$= -11 T is almost the same as the one at $B_z$= 11 T (Figure 2(c)), excluding the Zeeman splitting of degenerate LLs. Following the procedure used to extract the one group of LLs, we similarly label the first two peaks with the index $n = 4$ and adopt the phase factor γ=1/2 (Figure S4). Remarkably, the two groups of peaks converge into two TSS band dispersions (Figure 2(d)). In the energy range from -8 meV to 8 meV, the Fermi velocity of two dispersions remains about 1.4×10$^5$ m/s with band bottom at about -41.6 meV for band *a* and -42.1 meV for band *b*.

The multiple LLs is sensitive to the type of surface strain. In a region with strains different from those depicted in Figure 2 (Figure 3(a) and Figure S2(i-l)), three groups of LLs which are labeled as n$a$, n$b$ and n$c$ (n=0,1,2…) are observed in $B_z$-dependent d$I$/d$V$ spectra (Figure 3(b)) and Landau fan diagram (Figure 3(c)). Following the procedure before, we similarly label the first three peaks with the index $n = 4$ and adopt the phase factor γ=1/2 (Figure S5). Similarly, the three groups of peaks converge into three TSS band dispersions (Figure 3(d)). In the energy range from -8 meV to 8 meV, the Fermi velocity of three dispersions remains about 1.4×10$^5$ m/s, but the fitting band bottom are -41.0 meV for band $a$, -41.9 meV for band $b$ and -42.6 meV for band $c$. The three groups of LLs are homogeneous over the strained region (Fig. 3(e)).

To investigate the origin of the multiple LLs, we systematically study the strain effects based on DFT calculations and summarize our results in Figure 4. We firstly calculate the band structure of NiTe$_2$ (Figure S7(a)) and then track the evolution of the energy band belong to TSSs near E$_F$ under different types of strain. Based on the Fermi surfaces detected by the ARPES measurements[19], the TSSs bands form three pairs of electron pockets which are located at approximately midway site along three equivalent $\bar{\varGamma} - \bar{M}$ directions ($\bar{\varGamma} - \bar{M}_1$, $\bar{\varGamma} - \bar{M}_2$ and $\bar{\varGamma} - \bar{M}_3$), respectively (Figure 4(b)). In the region with uniform topographic height and without strain (Figure 4(a)), the electron-like subbands of TSS along the three directions are equivalent due to the crystalline symmetry, which results in one group of LLs (Figure 1(f) and Figure 4(b)). However, in the strained surface region, the crystalline symmetry is broken and the valley degeneration is lifted[30,31,33], which leads to the cascade of multiple groups of LLs. Under the uniaxial strain, such as strain along the crystalline zigzag (Figure 4(c)) or armchair direction (Figure 4(e)), there is an relative energy shift for one of the three subbands (Figure S7(c)). For example, when a strain up to 1% along the zigzag direction is applied, the TSS subband along the $\bar{\varGamma} - \bar{M}_3$ direction shifts -2.5 meV away from those along $\bar{\varGamma} - \bar{M}_1$ and $\bar{\varGamma} - \bar{M}_2$ directions (Figure 4(d)). Similarly, a strain of 1% along the armchair direction induces an energy shift of 1.8 meV between the subband along the $\bar{\varGamma} - \bar{M}_3$ direction (Figure 4(f)) and the other two directions. Accordingly, the valleys of the band under uniaxial strain produces two groups of LLs (right panel of Figure 4(d) and Figure 4(f)). The lattice distortion, two groups of LLs and their spectrum weights observed in the surface region of Figure 2 are consistent with the armchair strain case. Intriguingly, under the in-plane shear strain (Figure 4(g)), the 3-fold symmetry is fully broken and the valley degeneration along the three directions is completely lifted (Figure 4(h)). As a result, the minima of the three pairs of subbands are significantly different, inducing three sequences of LLs (right panel of Figure 4(h)). The lattice distortion, three groups of LLs and their spectrum weights observed in the surface region of Figure 3 are consistent with the shear strain case.

When considering the same energy splitting, the strain value in the FFT analysis (Figure S2) appears to be relatively larger than the calculated value (Figure 4). It is possibly caused by the inhomogeneous strain distribution present in the bulk $NiTe_2$. Specifically, the lattice distortions observed in the STM measurements correspond only to the top layer of bulk $NiTe_2$, whereas the DFT calculations assume a homogeneous distribution of strain in all layers of $NiTe_2$. It is therefore reasonable that the local lattice distortion at the surface of the top layer is larger than the calculated value due to the presence of inhomogeneities or large-sized surface wrinkles in the as-cleaved $NiTe_2$ sample in experiments. Nevertheless, the DFT calculations still qualitatively agree with the STM observations that multiple groups of LLs are indeed controlled by the types of strains (uniaxial or shear).

These findings reveal that the valley degeneracy of the TSSs can be effectively tuned by the different type of strain. However, it should be noted that STM is a local probe that can only detect the local strain present in the top layer of the $NiTe_2$ crystal. As such, it is difficult for us to experimentally estimate the magnitude of strain needed to obtain a given energy separation achievable with the state-of-the-art tunable strain apparatus. The quantitative relationship between the global strain and the separation of the TSSs remains for future investigations. The discovery of valley polarized TSSs in $NiTe_2$ reveals its potential for realizing a highly helicity-polarized[6], Josephson diode effect[24] and valley-polarized surface-isolated transport[34]. Since the formation of bulk Dirac cones and TSSs from a single orbital manifold is ubiquitous in TMDs[18], the lifting of the degenerate degrees of freedom via rotation symmetry breaking and the emergence of multigroup LLs can be extended to other TMDs such as $PdTe_2$[23], $PtSe_2$[35] and $WS_2$[36]. In addition, the monolayer $NiTe_2$[37] and doped $NiTe_2$[38,39] have been predicted to host superconducting ground state. Therefore, our findings will inspire further studies on the formation of superconducting TSSs[8,40] and topological superconductivity in the homojunctions/heterojunctions based on the TMDs thin films.

# FIGURES

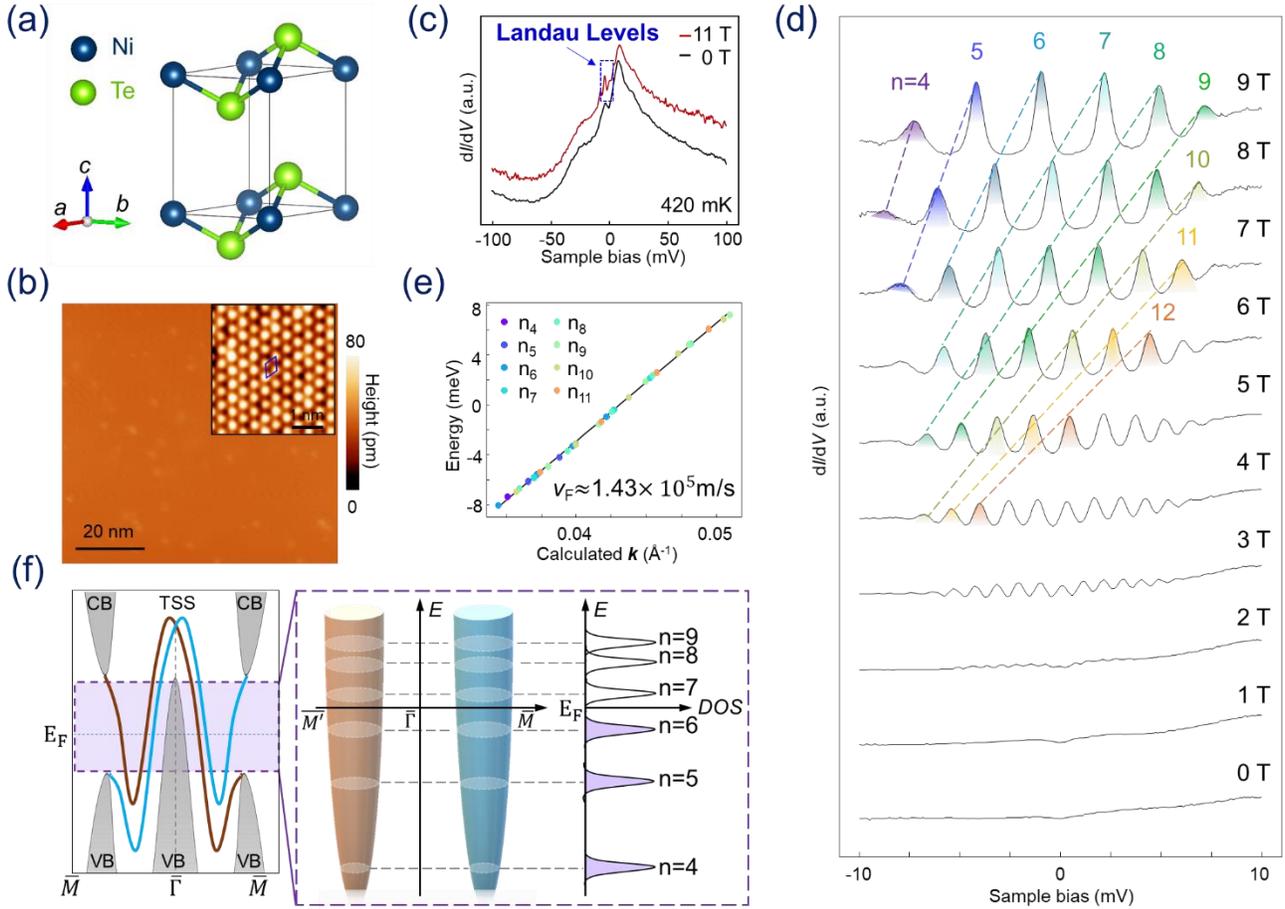

**Figure 1**. Landau quantization of the spin-orbit TSS band of the 1T-NiTe$_2$ crystal. (a), Crystal structures of layered 1T-NiTe$_2$. (b), Large-scale STM image, showing the flat and clean Te-terminated surface of NiTe$_2$ ($V_s$= -90 mV, $I_t$= 50 pA). Inset: Atomically-resolved STM image showing the hexagonal lattice of Te-terminated surface ($V_s$= -5 mV, $I_t$= 1 nA). (c), The spatially-averaged d$I$/d$V$ spectra obtained in the region (b) at the magnetic fields of 11 T and 0 T, respectively, showing that several peaks are emerge around the Fermi level at the magnetic field of 11 T ($V_s$= -100 mV, $I_t$= 7 nA, $V_{mod}$ = 1 mV). (d), Magnetic field dependent d$I$/d$V$ spectra obtained in the region (b), showing the evolution of Landau levels ($V_s$= -20 mV, $I_t$= 1 nA, $V_{mod}$ = 0.3 mV). (e), Calculated dispersion of topological surface states using the semiclassical Lifshitz-Onsager relation based on the Landau levels observed in (d). (f) Schematic showing spin-orbital polarized TSSs along the $\bar{\Gamma} - \bar{M}$ direction near the Fermi level, deduced from ARPES measurements and first-principles calculations[19,21] showing that the quantization of TSS produces Landau levels at a finite energy range. Each spectrum in (d) is vertically displaced for clarity.

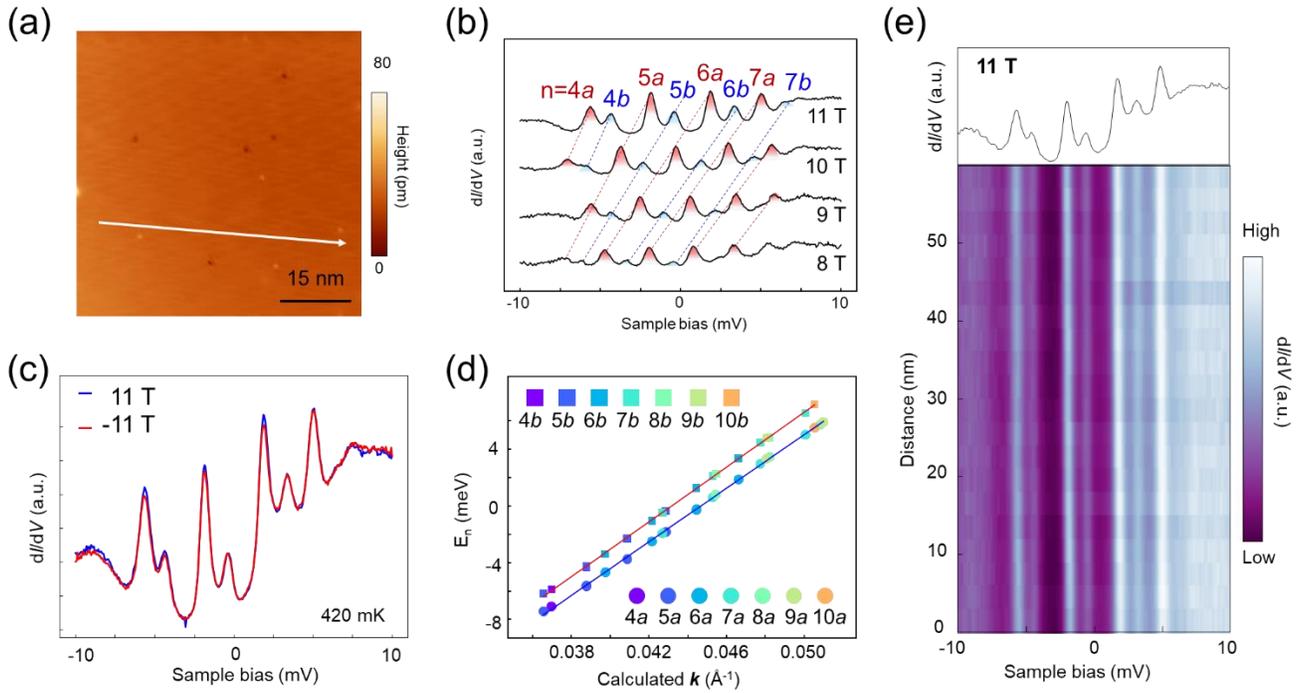

**Figure 2.** Two groups of Landau levels at a strained surface region. (a), STM image of a strained NiTe$_2$ surface region ($V_s$ = -90 mV, $I_t$ = 30 pA). (b), Magnetic field dependent d$I$/d$V$ spectra obtained in the region (a), showing the evolution of Landau levels. There are two groups of Landau levels highlighted by the red and blue colors, respectively ($V_s$ = -20 mV, $I_t$ = 1 nA, $V_{mod}$ = 0.3 mV). (c), d$I$/d$V$ spectra at the magnetic field of +11 and -11 T, respectively, showing that there is no difference in the intensity and energy positions of Landau levels under opposite fields ($V_s$ = -20 mV, $I_t$ = 1 nA, $V_{mod}$ = 0.3 mV). (d), Calculated dispersion of topological surface states using the semiclassical Lifshitz-Onsager relation based on the two groups of Landau levels observed in (b). (e), d$I$/d$V$ spectra along the white arrow in (a), showing that the Landau levels remain spatially homogeneous. The d$I$/d$V$ spectrum in the top panel is the first spectrum of the linecut ($V_s$ = -20 mV, $I_t$ = 1 nA, $V_{mod}$ = 0.3 mV). Each spectrum in (b) is vertically displaced for clarity.

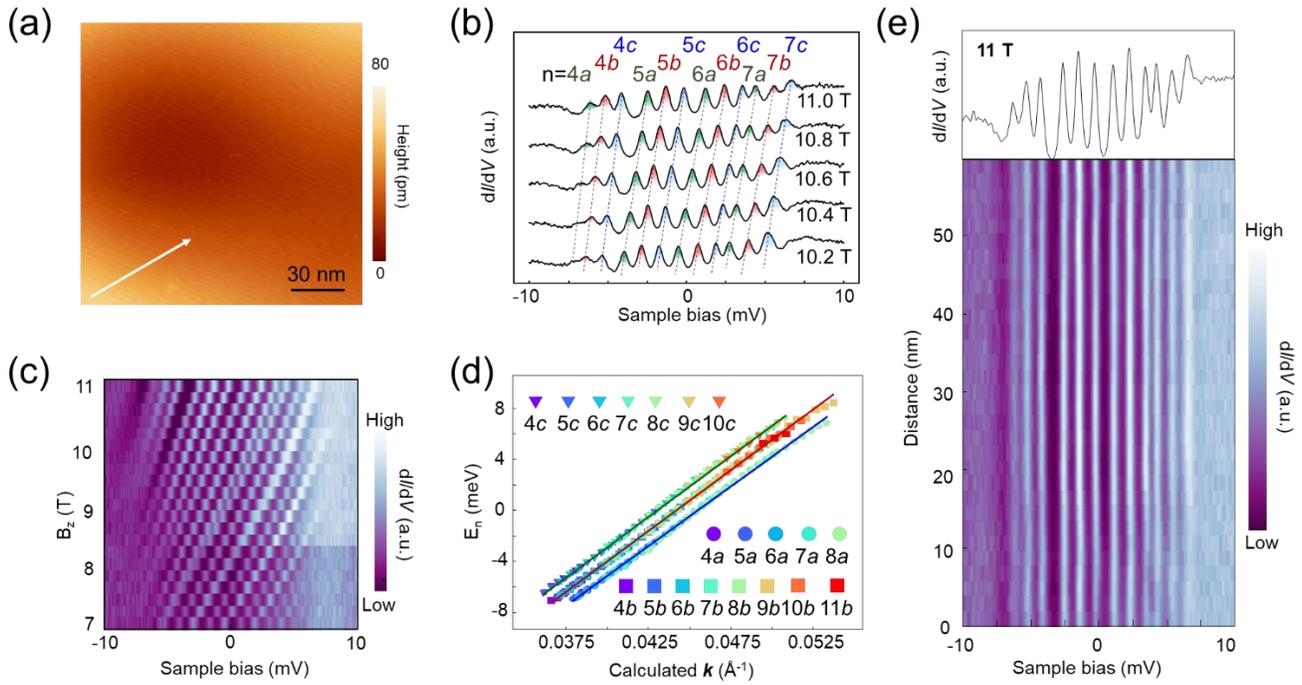

**Figure 3.** Three sequences of Landau levels at another strained surface region. (a), STM image of the strained surface region ($V_s$ = -90 mV, $I_t$ = 30 pA). (b), Magnetic field dependent d$I$/d$V$ spectra obtained in the region (a), showing three groups of Landau levels highlighted by the green, red and blue colors, respectively ($V_s$ = -10 mV, $I_t$ = 1.5 nA, $V_{mod}$ = 0.1 mV). (c), Landau fan diagram of (b) showing clearly that three groups of Landau levels shift to the higher energy with increasing magnetic fields ($V_s$ = -10 mV, $I_t$ = 1.5 nA, $V_{mod}$ = 0.1 mV). (d), Calculated dispersion of topological surface states using the semiclassical Lifshitz-Onsager relation based on three groups of Landau levels in (b). (e), d$I$/d$V$ spectra along the white arrow in (a), showing that the three groups of Landau levels are spatially homogeneous. The d$I$/d$V$ spectrum in the top panel is the first spectrum of the linecut ($V_s$ = -10 mV, $I_t$ = 1.5 nA, $V_{mod}$ = 0.1 mV). Each spectrum in (b) is vertically displaced for clarity.

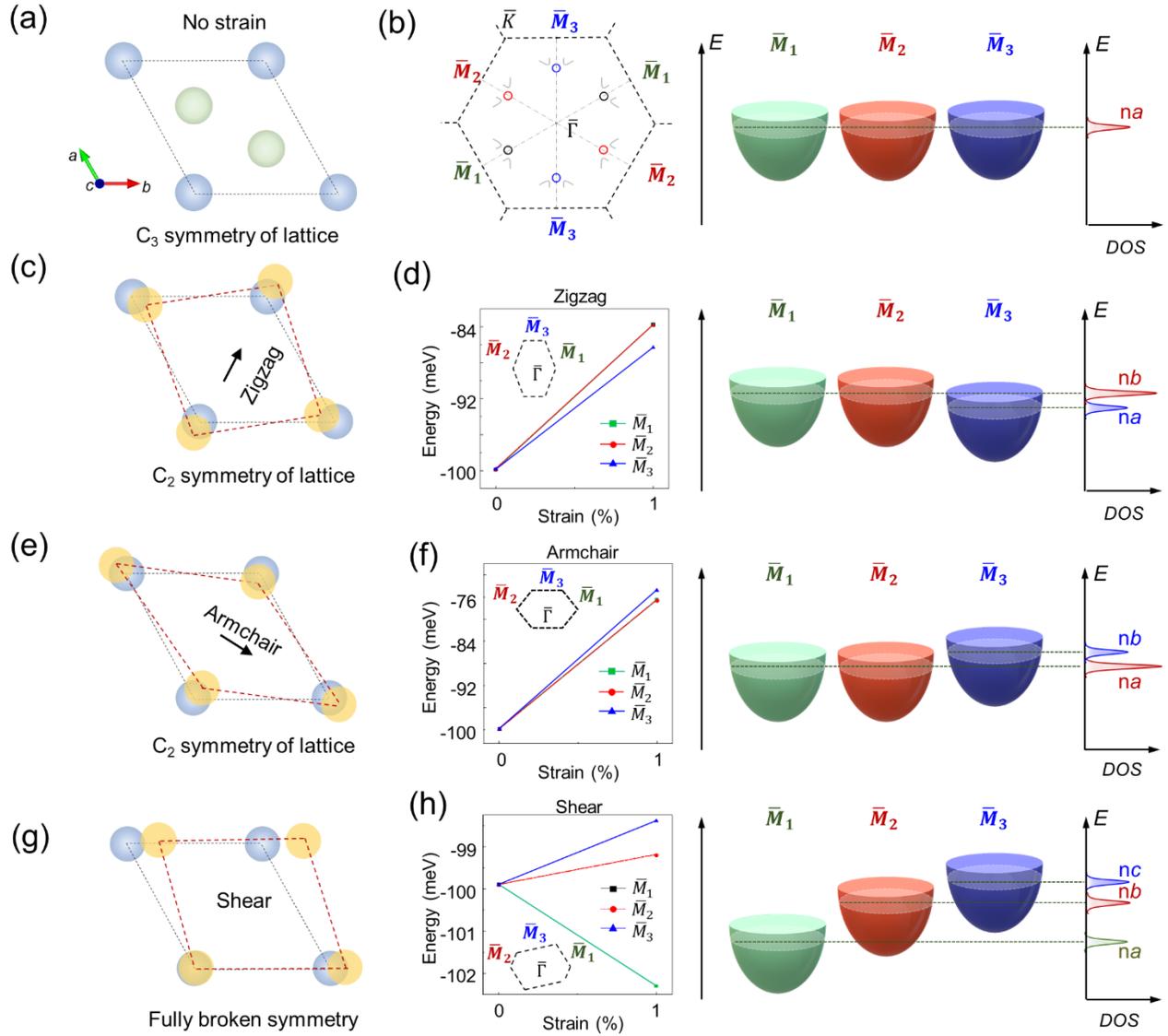

**Figure 4.** Tuning the multiple groups of Landau levels with distinct types of strain. (a), Schematic of the surface atomic structure of NiTe$_2$ without strain. (b), Left panel: Brillouin zone and Fermi surfaces of undistorted NiTe$_2$ deduced from ARPES measurements and first-principles calculations[19] showing the multivalleys of electron pockets of TSS branches. Right panel: Schematic showing the one group of Landau quantization of the TSSs in the regions without strains at the magnetic field. (c), Schematic of the surface atomic structure of NiTe$_2$ with zigzag strain. The distortion of the lattice is highlighted by the yellow balls and red dotted lines. (d) Left panel: The evolution of TSSs band bottom position of three symmetry directions $\bar{\Gamma}-\bar{M}_1$, $\bar{\Gamma}-\bar{M}_2$, and $\bar{\Gamma}-\bar{M}_3$ under zigzag strain calculated by DFT. Inset: Schematic of strain-induced deformation. Right panels: Schematic showing the two groups of Landau quantization of the TSSs in the regions with zigzag strains at the magnetic field. (e), Schematic of the surface atomic structure of NiTe$_2$ with armchair strain. The distortion of the lattice is highlighted by the yellow balls

and red dotted lines. (f) Left panel: The evolution of TSSs band bottom position of three symmetry directions $\bar{\Gamma}-\bar{M}_1$, $\bar{\Gamma}-\bar{M}_2$, and $\bar{\Gamma}-\bar{M}_3$ under armchair strain calculated by DFT. Inset: Schematic of strain-induced deformation. Right panel: Schematic showing the two groups of Landau quantization of the TSSs in the regions with armchair strains at the magnetic field. (g), Schematic of the surface atomic structure of NiTe$_2$ with shear strain. The distortion of the lattice is highlighted by the yellow balls and red dotted lines. (h), Left panel: The evolution of TSSs band bottom position of three symmetry directions $\bar{\Gamma}-\bar{M}_1$, $\bar{\Gamma}-\bar{M}_2$, and $\bar{\Gamma}-\bar{M}_3$ under shear strain calculated by DFT. Inset: Schematic of strain-induced deformation. Right panels: Schematic showing the three groups of Landau quantization of the TSSs in the regions with shear strains at the magnetic field.

# EXPERIMENTAL METHOD

**Single crystal growth and characterization of NiTe₂ sample**. Single crystals of NiTe₂ are synthesized *via* the self-flux method. The as-grown NiTe₂ single crystals were characterized by optical microscope (Olympus BX51-SC30). XRD pattern was collected using a Rigaku SmartLab SE X-ray diffractometer with Cu Kα radiation (λ=0.15418 nm) at room temperature. The rocking curve was achieved on BRUKER D8 VENTURE diffractometer with Mo Kα radiation (λ=0.71073 Å). The Raman spectrum was obtained by using Labram HR EV0, equipped with a diode laser emitting at 532 nm at a nominal power of 12 mW. The samples used in the TEM measurements were transferred onto a lacey support film after ground. The HRTEM image and SAED patterns were obtained by JEOL JEM-2100Plus microscope, operated at 200 kV.

**Scanning tunneling microscopy/spectroscopy.** The samples used in the STM/S experiments are cleaved at low temperature (78 K) and immediately transferred to an STM chamber. Experiments were performed in an ultrahigh vacuum (1×10⁻¹⁰ mbar) ultra-low temperature STM system equipped with 11 T magnetic field. All the scanning parameters (setpoint voltage and current) of the STM topographic images are listed in the figure captions. The electronic temperature in the low-temperature STS is 620 mK, calibrated using a standard superconductor, Nb crystal (Figure S8). Unless otherwise noted, the d*I*/d*V* spectra were acquired by a standard lock-in amplifier at a modulation frequency of 973.1 Hz. Non-superconducting tungsten tips were fabricated via electrochemical etching and calibrated on a clean Au(111) surface prepared by repeated cycles of sputtering with argon ions and annealing at 500 °C.

**Strain analysis and data fitting.** We analyze the strain distortions by comparing the length of three pairs of Bragg peaks in the Fourier transform of atomically-resolved STM images in strained regions (Figure S2)[41].

We assign the LL index *n* based on the semi-classical Lifshitz-Onsager relation. The extremal area $S_n$ in the reciprocal space of the *n*th Landau level at energy $E_n$ is given by $S_n = 2\pi e(n + \gamma)B/\hbar$, where γ is the phase factor. In the reciprocal space, we have $S_n = \pi k_n^2$, where $k_n$ is the geometric mean of the quantized orbital radius of $S_n$.

To obtain the accurate energy of the Landau peaks, we firstly substrate the Landau level spectroscopy by the spectrum with zero magnetic field. Then we use Lorentzian function:

$$y = y_0 + \frac{2A}{\pi} \frac{w}{4(x-x_c)^2 + w^2}$$

to fit the Landau peaks and extract the peak center $x_c$. Our fitting results match the original data very well (Figure S9) which makes the peak energy we get accurate.

**DFT calculations.** The structural optimizations and the electronic properties of layered NiTe$_2$ thin film were calculated in terms of the density functional theory (DFT)[42,43], with the projector augmented wave (PAW) potentials[44] as implemented in the Vienna Ab initio Simulation Package (VASP)[45]. The Perdew-Burke-Enzerhof approximation to the generalized-gradient approximation (GGA)[46] is adopted. Our computational slab model consists of ten NiTe$_2$ unit-cell layers, and a vacuum layer with thickness of 18 Å. For structural optimization calculations, we use a 14×14×1 Gamma-center k-point mesh and an energy cutoff 550 eV for convergence. The lattice constants are relaxed by minimizing the total energies to an accuracy of 0.001 meV per primitive unit cell, and the ion positions are relaxed with an accuracy of 0.001 eV/Å. The scalar relativistic approximation is used to treat the relativistic effects. The spin-orbit coupling is also included in calculating the band structures of this compounds.